\documentclass[12pt,preprint]{aastex}
\begin{document}

\title{Mapping Observations of DNC and HN$^{13}$C in Dark Cloud Cores}
\author{Tomoya HIROTA\altaffilmark{1}}
\affil {Department of Physics, Faculty of Science, 
Kagoshima University,}
\affil {Korimoto 1-21-35, Kagoshima 890-0065, JAPAN; 
hirota@sci.kagoshima-u.ac.jp}
\altaffiltext{1}{Current address: Earth Rotation Division, 
National Astronomical Observatory of Japan, 
Osawa 2-21-1, Mitaka, Tokyo 181-8588, Japan; 
tomoya.hirota@nao.ac.jp
}
\author{Masafumi IKEDA}
\affil {RIKEN(The Institute of Physical and Chemical Research),}
\affil {Hirosawa 2-1, Wako, Saitama 351-0198, JAPAN}
\author{Satoshi YAMAMOTO}
\affil {Department of Physics and 
Research Center for the Early Universe,}
\affil {The University of Tokyo, Bunkyo-ku, Tokyo 113-0033, JAPAN }

\begin{abstract}
We present results of mapping observations of 
the DNC, HN$^{13}$C, and H$^{13}$CO$^{+}$ lines ($J$=1-0) 
toward 4 nearby dark cloud cores, TMC-1, L1512, L1544, and L63, 
along with observations of the DNC and HN$^{13}$C lines ($J$=2-1) 
toward selected positions. 
By use of statistical equilibrium calculations based on the 
LVG model, the H$_{2}$ densities are derived to be 
(1.4-5.5)$\times$10$^{5}$ cm$^{-3}$, and 
the [DNC]/[HN$^{13}$C] ratios are derived to be 1.25-5.44 
with a typical uncertainty by a factor of 2. 
The observed [DNC]/[HNC] ratios range from 0.02 to 0.09, 
assuming the [$^{12}$C]/[$^{13}$C] ratio of 60. 
Distributions of DNC and HN$^{13}$C are generally 
similar to each other, whereas 
the distribution of H$^{13}$CO$^{+}$ is more extended than 
those of DNC and HN$^{13}$C, indicating that they reside in an 
inner part of the 
cores than HCO$^{+}$. 
The [DNC]/[HN$^{13}$C] ratio is rather constant within each core, 
although a small systematic gradients are observed in 
TMC-1 and L63. Particularly, no such systematic gradient is found 
in L1512 and L1544, 
where a significant effect of depletion of molecules is reported 
toward the central part of the cores. 
This suggests that the [DNC]/[HNC] ratio would not be very 
sensitive to depletion factor, 
unlike the [DCO$^{+}$]/[HCO$^{+}$] ratio. On the other hand, 
the core to core variation of 
the [DNC]/[HNC] ratio, which range an order of 
magnitude, is more remarkable than the variation within each core. 
These results are interpreted qualitatively by 
a combination of three competing time-dependent processes; 
gas-phase deuterium fractionation, depletion of molecules onto 
grain surface, and dynamical evolution of a core. 
\end{abstract}


\section{Introduction}

It has been recognized that the deuterium 
atom is fractionated into various molecules in cold interstellar 
clouds and the isotopic abundance ratios of the deuterated molecules 
relative to their normal species are as high as 10\%, although the 
cosmic [D]/[H] ratio is only an order of 10$^{-5}$ 
(Wilson \& Rood 1994). 
Deuterium fractionation is thought to be caused by the following 
exothermic isotope-exchange reaction: 
\begin{eqnarray}
\mbox{HX}^{+} + \mbox{HD} & \rightarrow & 
\mbox{DX}^{+} + \mbox{H}_{2} + \Delta E, 
\label{eq-hxp}
\end{eqnarray}
where HX$^{+}$ represents a molecular ion such as H$_{3}^{+}$ 
and H$_{3}$O$^{+}$ 
(Millar, Bennet, \& Herbst 1989; Howe \& Millar 1993). 

Since the deuterium fractionation processes are sensitive to the 
physical conditions such as temperature, ionization degree, and 
depletion factor of molecules, 
deuterated molecules can be used as unique tracers 
to investigate physics and chemistry of dense cores in 
dark clouds. 
Survey observations of various deuterated molecules 
such as DCO$^{+}$ (Butner, Lada, \& Loren 1995), 
C$_{3}$HD (Gerin et al. 1987; Bell et al. 1988), 
DC$_{3}$N (Howe et al. 1994), 
NH$_{2}$D (Saito et al. 2000; Shah \& Wootten 2001), 
and DNC (Hirota, Ikeda, \& Yamamoto 2001, hereafter Paper I) 
have been carried out toward 
a number of dense cores. 
Furthermore, various model calculations 
have been presented in order to account for the observed deuterium 
fractionation (Millar et al. 1989; Howe \& Millar 1993; 
Roberts \& Millar 2000). 
A detailed study on chemistry of deuterated molecules has recently 
been carried out with sensitive observations of 10 
deuterated molecules as well as chemical model calculations 
(Turner 2001). 
According to these observations, 
deuterium fractionation of various molecules significantly varies 
from core to core, and also within each core, which 
is interpreted in terms of the difference in degree of depletion, 
ionization fraction, chemical bistability, and/or chemical evolution
(Turner 2001 and references therein). 
However, detailed observations are still limited to a small 
number of sources such as TMC-1, L1544, and L183, and also to a few 
specific molecular species such as DCO$^{+}$, N$_{2}$D$^{+}$, and 
NH$_{2}$D. 

In order to study an origin of the variation of deuterium 
fractionation, it is important to investigate a possible relation 
between deuterium fractionation and abundances of other molecules. 
Such a comprehensive study 
would constrain mechanisms of the variation more strongly. 
On the basis of this motivation, we carried out survey observations 
of DNC and HN$^{13}$C toward 29 dark cloud cores, and found that 
deuterium fractionation of [DNC]/[HNC] 
(hereafter we just call DNC/HNC ratio) has a rough correlation with 
the NH$_{3}$/CCS ratio (Paper I). Since the NH$_{3}$/CCS ratio 
is regarded as an indicator of chemical evolution 
(Suzuki et al. 1992), this result suggests that deuterium 
fractionation reflects, at least to some extent, the chemical 
evolutionary stage of dark cloud cores. Particularly, a significant 
gradient in the DNC/HNC ratio is seen along the ridge of TMC-1, 
which also shows the correlation with the NH$_{3}$/CCS ratio. 
Saito et al. (2002) also pointed out the gradient in 
the DCO$^{+}$/HCO$^{+}$ ratio along the TMC-1 ridge, 
suggesting that it originates from the chemical evolutionary 
effect on the basis of chemical model calculations. 

Now it is of particular importance to study whether deuterium 
fractionation varies within a dense core, 
as observed in TMC-1, or not. 
With this in mind, we have carried out mapping observations of the 
$J$=1-0 lines of DNC, HN$^{13}$C, and H$^{13}$CO$^{+}$ toward 
4 nearby dark cloud cores, TMC-1, L1512, L1544, and L63. 
Furthermore, we have also conducted observations of 
the $J$=2-1 lines of DNC and HN$^{13}$C toward selected positions 
to derive their column densities and H$_{2}$ densities accurately. 
A related study for another dark cloud core, L1521E, 
has already been reported by Hirota, Ito, \& Yamamoto (2002). 

\section{Observations}

Mapping observations of the DNC, HN$^{13}$C, and H$^{13}$CO$^{+}$ 
lines ($J$=1-0) 
were made in several observing sessions from 1997 to 1999 with 
the 45 m radio telescope at Nobeyama Radio Observatory\footnotemark 
\footnotetext{Nobeyama Radio Observatory (NRO) is a branch of 
the National Astronomical Observatory of Japan, an interuniversity 
research institute operated by 
the Ministry of Education, Culture, Sports, 
Science and Technology of Japan}. 
Their rest frequencies, dipole moments, and intrinsic line strengths
are summarized in Table \ref{tab-obsedf}. 
The beam sizes (HPBW) were 20\arcsec \ and 17\arcsec \ for the 
76 GHz and 87 GHz regions, respectively. 
We used SIS mixer receivers whose system temperatures were typically 
300 K (SSB) including atmospheric attenuation. 
Toward L1512 and L63, the spectra of DNC, HN$^{13}$C, 
and H$^{13}$CO$^{+}$ were simultaneously observed 
with two different receivers by using a polarization beam splitter; 
one receiver was used for DNC (76 GHz), and the other for 
HN$^{13}$C and H$^{13}$CO$^{+}$ (87 GHz). 
The relative pointing error between these two receivers was 
less than 1\arcsec. 
Toward TMC-1 and L1544, the HN$^{13}$C and H$^{13}$CO$^{+}$ 
lines were observed simultaneously with the two receivers, while 
the DNC line was observed separately with only one receiver. 
The main beam brightness temperature is determined by dividing the 
observed antenna temperature by the main beam efficiency, 
$\eta_{mb}$, which was provided by the observatory to be 0.5 in 
the 76-87 GHz region. 
Acousto-optical radio spectrometers were used for the backend. 
The spectral resolution was 37 kHz, which corresponds to velocity 
resolutions of 
0.15 km s$^{-1}$ and 0.13 km s$^{-1}$ in the 76 GHz and 87 GHz 
regions, respectively. 
Pointing was checked by observing nearby SiO maser sources 
every 1-2 hours, and the pointing accuracy was 
estimated to be better than 5\arcsec \ (rms). 
All the observations were performed with the position-switching 
mode, in which a typical off position was 10\arcmin \ away 
from the source position. 

We observed 4 dark cloud cores, TMC-1, L1512, and L1544 in 
the Taurus region and L63 in the Ophiuchus region, 
as summarized in Table \ref{tab-mapobs}. 
The reference positions were taken from Suzuki et al. (1992) and 
Benson \& Myers (1989) for TMC-1 and other cores, respectively. 
Note that the reference position for L1544 listed in 
Table \ref{tab-mapobs} is shifted by 
(-27\arcsec,64\arcsec) from that commonly 
used, which is determined by the high-resolution 
observation of dust continuum emission 
(e.g. Ward-Thompson et al. 1994). 
The grid spacing (20\arcsec-60\arcsec) in the mapping observations 
was larger than the beam size, and hence, maps are undersampled. 
The on-source integration time ranged from 120 to 360 
seconds for each position, 
and the rms noise temperature ranged from 0.14 to 0.25 K in 
the $T_{mb}$ scale. 

In addition, the $J$=2-1 lines of DNC and HN$^{13}$C were observed 
toward selected positions with the NRAO 12 m telescope\footnotemark 
\footnotetext{The National Radio Astronomy Observatory (NRAO) is a 
facility of the National Science Foundation, operated under 
cooperative agreement by Associated Universities, Inc.}
at Kitt Peak in 1999 December. The beam sizes were 
40\arcsec \ and 35\arcsec  (HPBW) for the DNC and HN$^{13}$C 
observations, respectively. We used two SIS mixer 
receivers and observed the DNC and HN$^{13}$C lines 
simultaneously. The system temperatures were typically 400-500 K 
(SSB), although those for the HN$^{13}$C lines were systematically 
higher because of the heavier atmospheric absorption due to the 
water line (183 GHz). 
Since the temperature scale is given as $T_{R}^{*}$ in the 
observations with the NRAO 12 m telescope, we converted these values 
to the main beam brightness temperature by the following equation; 
$T_{mb}=T_{R}^{*}/\eta_{m}^{*}$. The corrected 
main beam efficiency ($\eta_{m}^{*}$) of the NRAO 12 m telescope was 
0.7 for the DNC and HN$^{13}$C lines ($J$=2-1), 
which was derived from observations of planets. 
The hybrid spectrometers with the frequency resolution of 
48.8 kHz were used for the backend. The velocity resolutions 
were 0.096 km s$^{-1}$ and 0.084 km s$^{-1}$ in 
the 152 GHz and 174 GHz regions, respectively. 
We employed the position-switching mode as in the observations 
with the NRO 45 m telescope. 
On source integration time was from 20 to 70 minutes and the rms 
noise temperature was from 0.04 to 0.15 K in the $T_{mb}$ scale. 

\section{Line Parameter}

\begin{figure}
\epsscale{0.9}
\plottwo{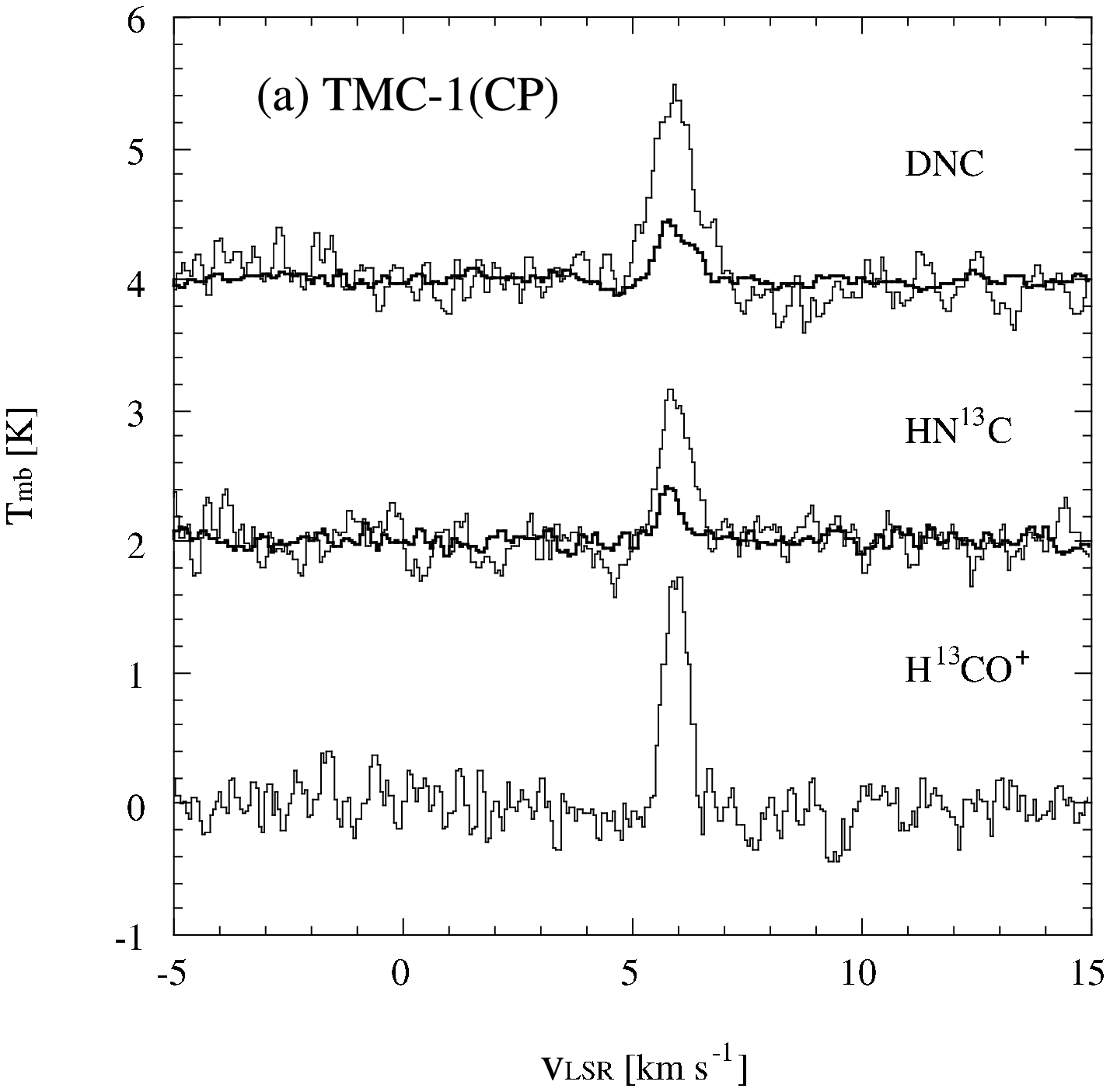}{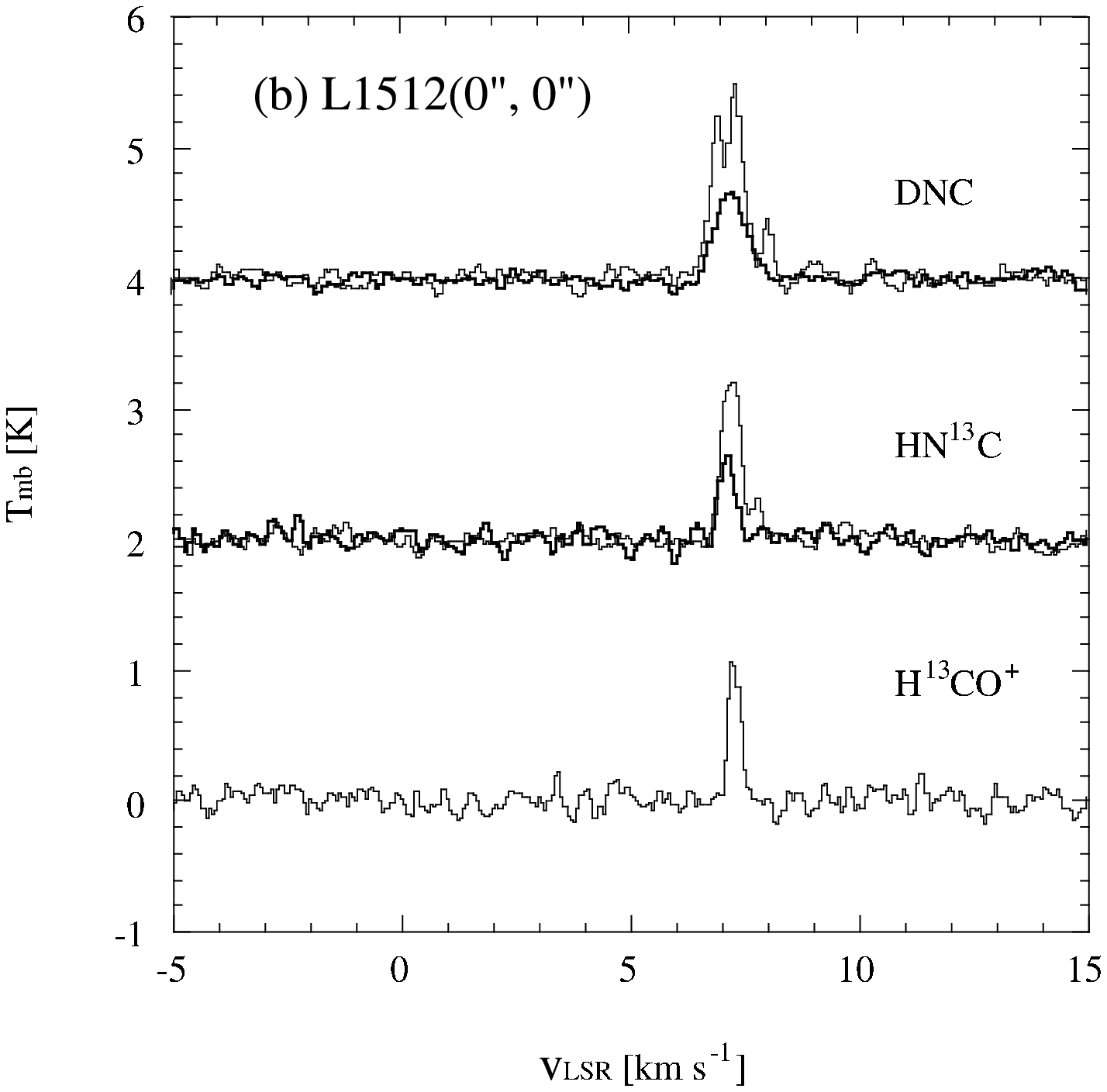}
\epsscale{2}
\plottwo{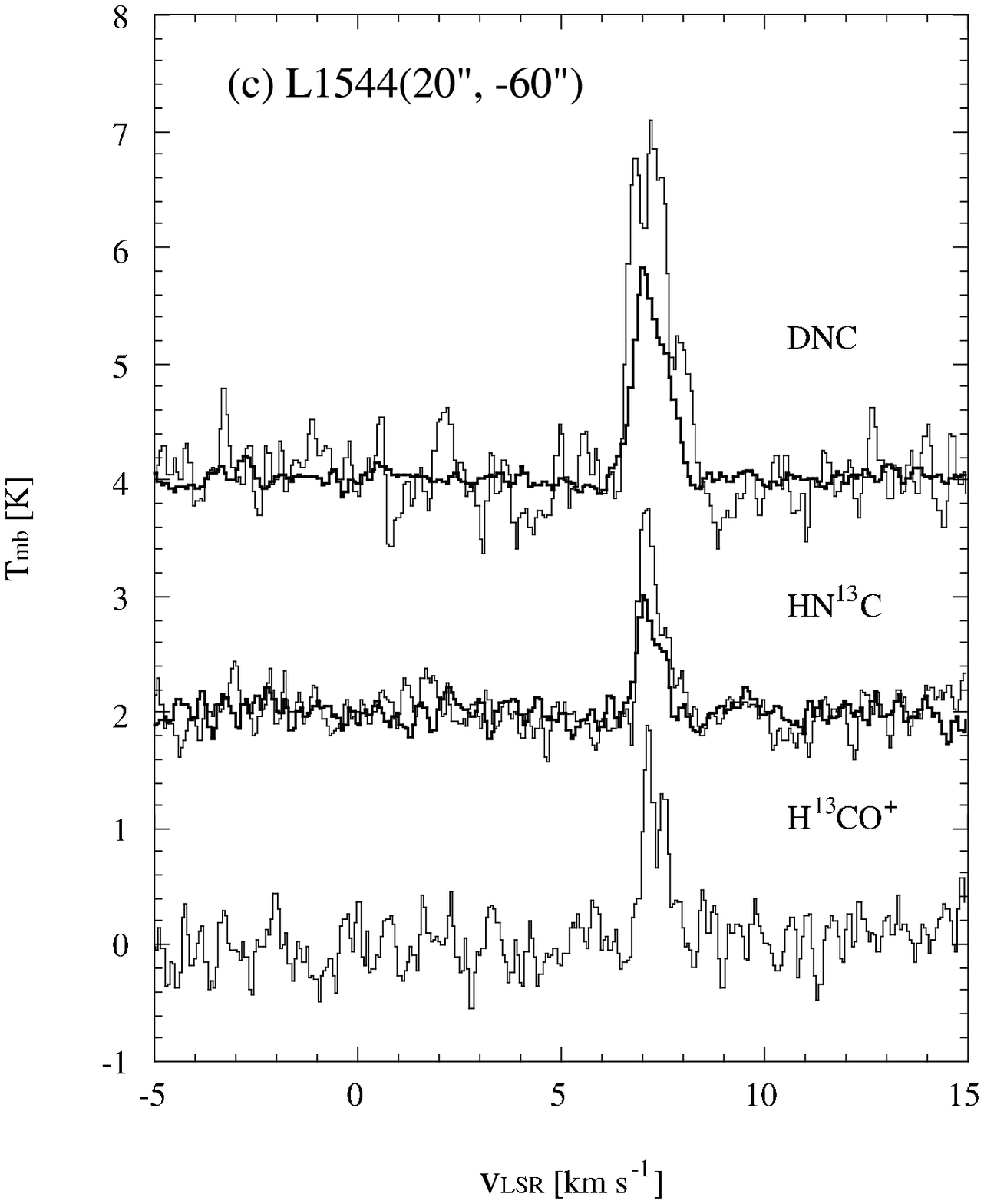}{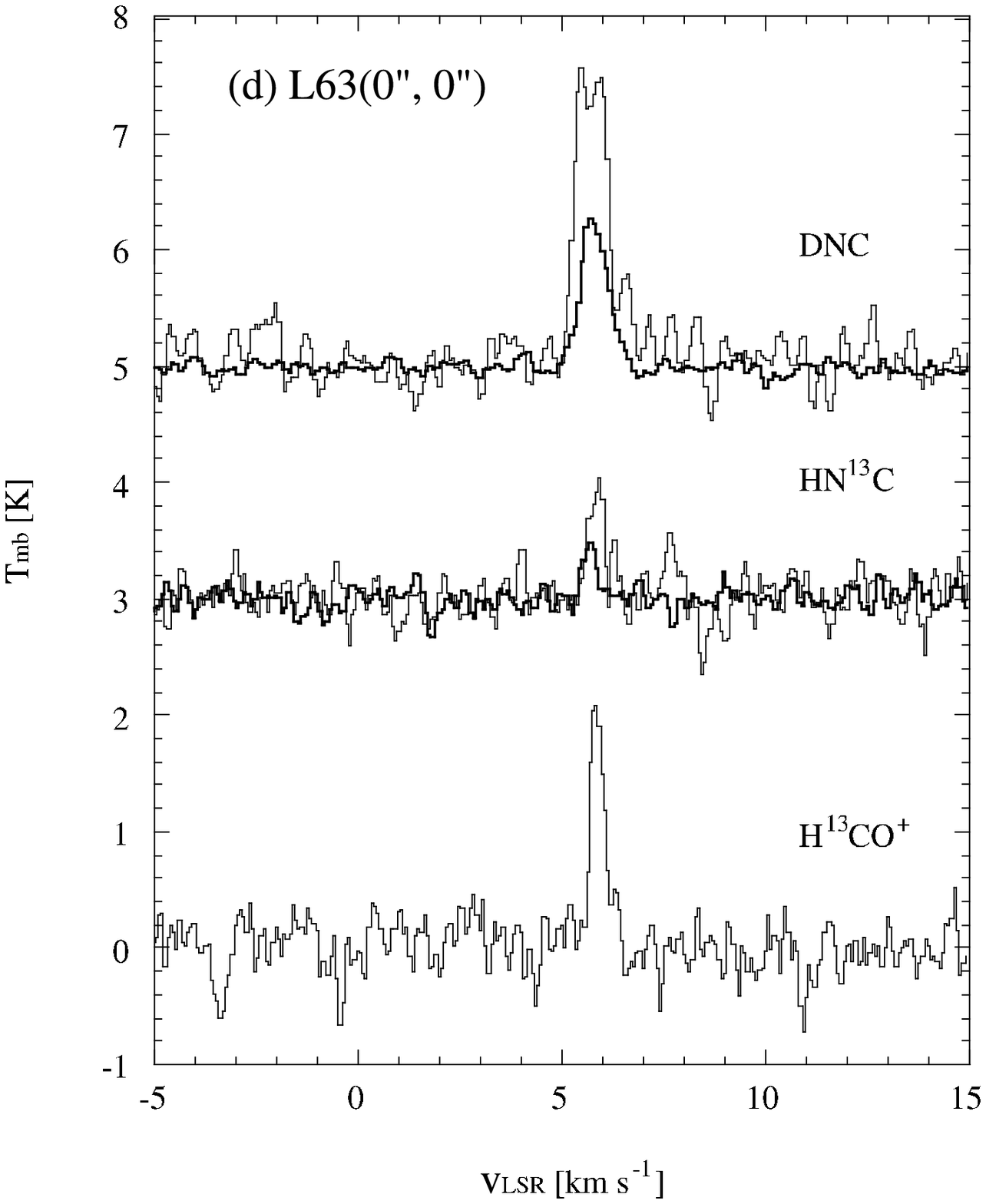}
\caption{Spectra of DNC, HN$^{13}$C, and 
H$^{13}$CO$^{+}$ toward the selected positions in 
the observed cores; (a)TMC-1(CP), 
(b)L1512, (c)L1544, and (d)L63. 
Thick and thin lines represent the $J$=2-1 and 1-0 lines, respectively. 
\label{fig-spmap} }
\end{figure}

We observed both the $J$=1-0 and 2-1 lines of DNC and HN$^{13}$C 
toward selected 10 positions. The DNC lines were detected toward 
all the positions, while the $J$=1-0 and 2-1 lines of HN$^{13}$C 
were detected toward 9 and 8 positions, respectively. 
Examples of the observed spectra are shown in Figure  \ref{fig-spmap}. 
It is known that the DNC and HN$^{13}$C lines have complex hyperfine 
structure due to the deuterium and nitrogen 
nuclei (Frerking, Langer, \& Wilson 1979; Paper I; Turner 2001). 
Since the hyperfine splittings in the DNC and HN$^{13}$C lines 
were not resolved completely due to the small nuclear quadrupole 
coupling constants, $eQq$ (Paper I; Turner 2001), 
a single Gaussian profile was fitted to each spectrum to determine 
the peak antenna temperature, line width, and LSR velocity. 
Results are summarized in Table \ref{tab-obs}. 

As shown in Table \ref{tab-obs}, 
broadening due to an unresolved hyperfine structure pattern is 
larger for the DNC lines than for the HN$^{13}$C lines, 
and is larger for the $J$=1-0 lines than for the $J$=2-1 lines. 
In addition, the LSR velocities of the $J$=1-0 and 2-1 lines of 
DNC and HN$^{13}$C seem to be slightly different even toward the 
same position, which are also due to the hyperfine structure of 
the lines. Therefore, the line parameters derived from 
the single Gaussian fitting do not exactly represent the velocity 
width of the core. Although it is, in principle, possible to 
determine the velocity width by the Gaussian fitting including 
hyperfine structure, 
it is almost impossible to do this because of the limited 
spectral resolution of the spectrometers used in our observations. 
In practice, the derived column densities are consistent with those 
derived by including 
the hyperfine splittings, as shown in Paper I and Turner (2001). 
Thus, we analyzed our data using the results of the single 
Gaussian fitting. 

We found a double-peaked line profile in 
the H$^{13}$CO$^{+}$ spectrum toward L1544, indicating that 
there are two velocity components in L1544 
(Tafalla et al. 1998; Caselli et al. 2002a). 
On the other hand, the DNC and HN$^{13}$C lines do not show 
apparent evidence for two velocity components or a self-absorption 
dip, indicating the difference in distributions between HCO$^{+}$ 
and HNC; HNC resides in an inner part of the core while 
HCO$^{+}$ tend to be depleted at the core center. 

\section{LVG Analysis}

We analyzed the $J$=2-1 and 1-0 data of DNC and HN$^{13}$C 
by statistical equilibrium calculations based on the 
large velocity gradient (LVG) model (Goldreich \& Kwan 1974). 
We simultaneously fitted the DNC and HN$^{13}$C data to derive 
the H$_{2}$ density and the column densities of DNC and HN$^{13}$C. 
Since the hyperfine structure was not resolved for 
most of the observed lines, we ignored the hyperfine splitting. 
Collisional excitation rates for the DNC-H$_{2}$ system 
and the HN$^{13}$C-H$_{2}$ system were calculated from those 
of  the HCN-He system (Green \& Thaddeus 1974). 
We multiplied by a factor of 1.37 to the HCN-He collision rates to 
correct the difference in the reduced mass. 
The dipole moment is 3.05 debye both for DNC and HN$^{13}$C 
(Blackman et al. 1976). 

The mean FWHM line width of the $J$=2-1 and 1-0 lines 
was used to estimate the velocity gradient in the LVG calculations. 
This assumption would affect the derived column density, because 
the $J$=2-1 lines have systematically narrower line widths 
than the $J$=1-0 lines due to a difference in the 
unresolved hyperfine structure. However, we confirmed that the 
systematic errors in the column density caused by this assumption 
would be less than 25\%. 
The kinetic temperature was 
assumed to be 10 K for all the cores (Benson \& Myers 1989). 
If we assume the kinetic temperature of 8-12 K, the derived column 
densities of DNC and HN$^{13}$C vary only 10-20 \%, and hence, 
a change in the DNC/HN$^{13}$C ratios is confirmed to be less 
than 20 \%, while the H$_{2}$ densities typically vary by 
a factor of 2. 

The H$_{2}$ densities and the column densities of DNC and HN$^{13}$C 
obtained by the above method are summarized in Table \ref{tab-lvg}, 
and the derived DNC/HN$^{13}$C ratios are summarized in 
Table \ref{tab-dhratio}. 
In this analysis, we noticed that the intensities of the $J$=2-1 
lines relative to the $J$=1-0 lines of DNC observed in 
TMC-1(cyanopolyyne peak, hereafter CP), 
L63(-30\arcsec, 90\arcsec), and L63(0\arcsec, 60\arcsec) are 
slightly smaller than those allowed by the LVG model. 
This is in part due to the limited signal-to-noise ratio of 
the data, but may also reflect a systematic error such that 
the DNC lines are more affected by the beam dilution effect 
than the HN$^{13}$C lines because of a systematically larger 
beam size for DNC. 
Therefore, we assumed the H$_{2}$ density derived from the 
HN$^{13}$C data in the analysis for TMC-1 (CP). 
Similarly, we employed the H$_{2}$ density derived from the DNC 
and HN$^{13}$C data toward the (0\arcsec, 0\arcsec) position of 
L63 in the analysis for the other 2 positions in L63. 

For comparison, we analyzed the DNC and HN$^{13}$C data individually. 
We found that the H$_{2}$ densities derived from the HN$^{13}$C data 
are larger by a factor of 1.3-4.5 than those from the DNC data. 
As a result, 
the DNC/HN$^{13}$C ratios systematically increased by a factor of 
1.2-2.6. 
The largest change in the DNC/HN$^{13}$C ratios were found for 
the (0\arcsec, 0\arcsec) position of L63 and TMC-1 (core C), 
where the change in the DNC/HN$^{13}$C ratios are factor of 
2.6 and 2.4, respectively. 
Except for these two positions, the uncertainty in the DNC/HN$^{13}$C 
ratio due to the assumed H$_{2}$ density 
is estimated to be less than a factor of 1.6. 
In total, the typical uncertainty 
in the DNC/HN$^{13}$C ratio is estimated to be a factor of 2. 

The H$_{2}$ densities determined by the LVG analysis 
range from 1.4$\times$10$^{5}$ cm$^{-3}$ to 
5.5$\times$10$^{5}$ cm$^{-3}$. 
These values indicate that the DNC and HN$^{13}$C molecules 
seem to be distributed in similar density regions traced by 
DCO$^{+}$ (Butner et al. 1995), C$^{34}$S (Hirota et al. 1998), 
NH$_{3}$ (Benson \& Myers 1989), and N$_{2}$H$^{+}$ 
(Caselli et al. 2002c). 
The density variation is not very large among the observed positions. 
We compared our results with those of 
Paper I, in which the H$_{2}$ densities determined from 
the C$^{34}$S lines were used (Hirota et al. 1998). 
For three positions, TMC-1(CP), TMC-1(NH$_{3}$), and 
L63 (0\arcsec, 0\arcsec), we found that the H$_{2}$ densities 
derived from the DNC and HN$^{13}$C lines are comparable to those 
derived from the C$^{34}$S lines (Hirota et al. 1998). 
On the other hand, the H$_{2}$ density determined from 
the C$^{34}$S lines for L1544 (0\arcsec, 0\arcsec) is lower 
by a factor of 5 than that from the DNC and HN$^{13}$C lines. 
Since the (0\arcsec, 0\arcsec) position of L1544 is not 
the core center but rather close to the edge, the H$_{2}$ density 
derived from the C$^{34}$S data would be reduced. 
Nevertheless, the DNC/HN$^{13}$C ratios reported for these positions 
in Paper I are comparable to those derived in the present study. 

The derived DNC/HN$^{13}$C ratios range from 1.25 to 5.44. 
The DNC/HN$^{13}$C ratio for TMC-1(CP), 1.25,  agrees well with 
that of Turner (2001), 0.99-1.24. 
Assuming that the $^{12}$C/$^{13}$C ratio is 60 
(Mangum et al. 1988; Langer \& Penzias 1993; 
Ikeda, Hirota, \& Yamamoto 2002), the DNC/HNC ratios are derived 
to be 0.02-0.09. 
These values are consistent with our previous estimates 
(Paper I). 

\section{Distributions of DNC and HN$^{13}$C}

\begin{figure}
\epsscale{1}
\plotone{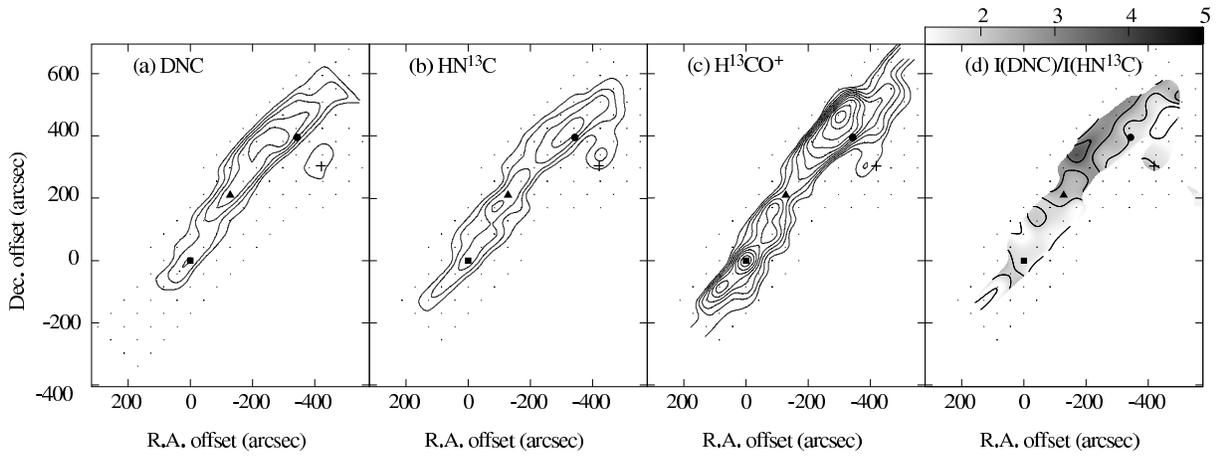}
\caption{Integrated intensity maps of TMC-1. 
Filled square, filled circle, and filled triangle represent the 
positions of cyanopolyyne peak, ammonia peak, and core C, 
respectively. The cross represents 
the IRAS source (IRAS 04381+2540). 
(a) DNC ($J$=1-0). 
The velocity range of integration is from 4.5 to 7.0 km s$^{-1}$. 
The interval of the contours is 0.30 K km s$^{-1}$ and 
the lowest one is 0.60 K km s$^{-1}$. 
(b) HN$^{13}$C ($J$=1-0). 
The velocity range of integration is from 4.6 to 6.8 km s$^{-1}$. 
The interval of the contours is 0.16 K km s$^{-1}$ and 
the lowest one is 0.32 K km s$^{-1}$. 
(c) H$^{13}$CO$^{+}$ ($J$=1-0). 
The velocity range of integration is from 5.0 to 6.6 km s$^{-1}$. 
The interval of the contours is 0.10 K km s$^{-1}$ and 
the lowest one is 0.60 K km s$^{-1}$.  
(d) Integrated intensity ratio of DNC/HN$^{13}$C ($J$=1-0). 
Contour levels are 10, 30, 50, 70, and 90 \% of the peak value (3.72). 
\label{fig-tmc1map} }
\end{figure}

\begin{figure}
\epsscale{1}
\plotone{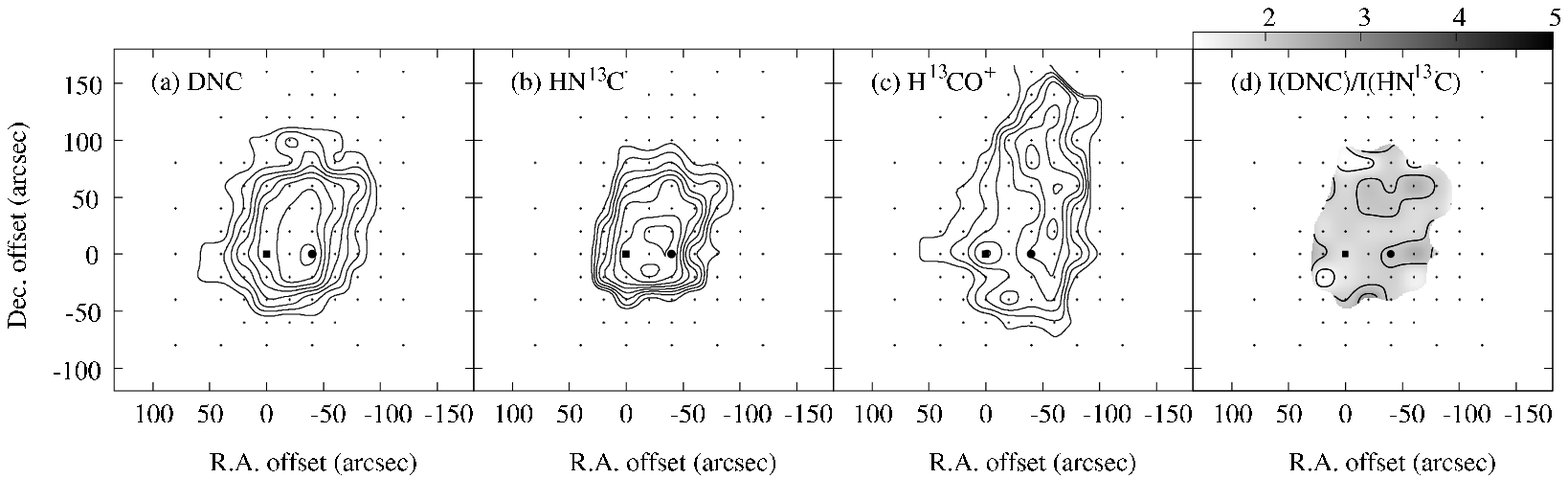}
\caption{Integrated intensity maps of L1512. 
Filled square and filled circle represent the (0\arcsec,0\arcsec) 
position and (-40\arcsec, 0\arcsec) position, respectively.  
(a) DNC ($J$=1-0). 
The velocity range of integration is from 6.5 to 7.6 km s$^{-1}$. 
The interval of the contours is 0.12 K km s$^{-1}$ and 
the lowest one is 0.24 K km s$^{-1}$. 
(b) HN$^{13}$C ($J$=1-0). 
The velocity range of integration is from 6.7 to 7.5 km s$^{-1}$. 
The interval of the contours is 0.05 K km s$^{-1}$ and 
the lowest one is 0.20 K km s$^{-1}$. 
(c) H$^{13}$CO$^{+}$ ($J$=1-0). 
The velocity range of integration is from 6.8 to 7.5 km s$^{-1}$. 
The interval of the contours is 0.05 K km s$^{-1}$ and 
the lowest one is 0.20 K km s$^{-1}$. 
(d) Integrated intensity ratio of DNC/HN$^{13}$C ($J$=1-0). 
Contour levels are 10, 30, 50, 70, and 90 \% of the peak value (2.86). 
\label{fig-l1512map} }
\end{figure}

\begin{figure}
\epsscale{1}
\plotone{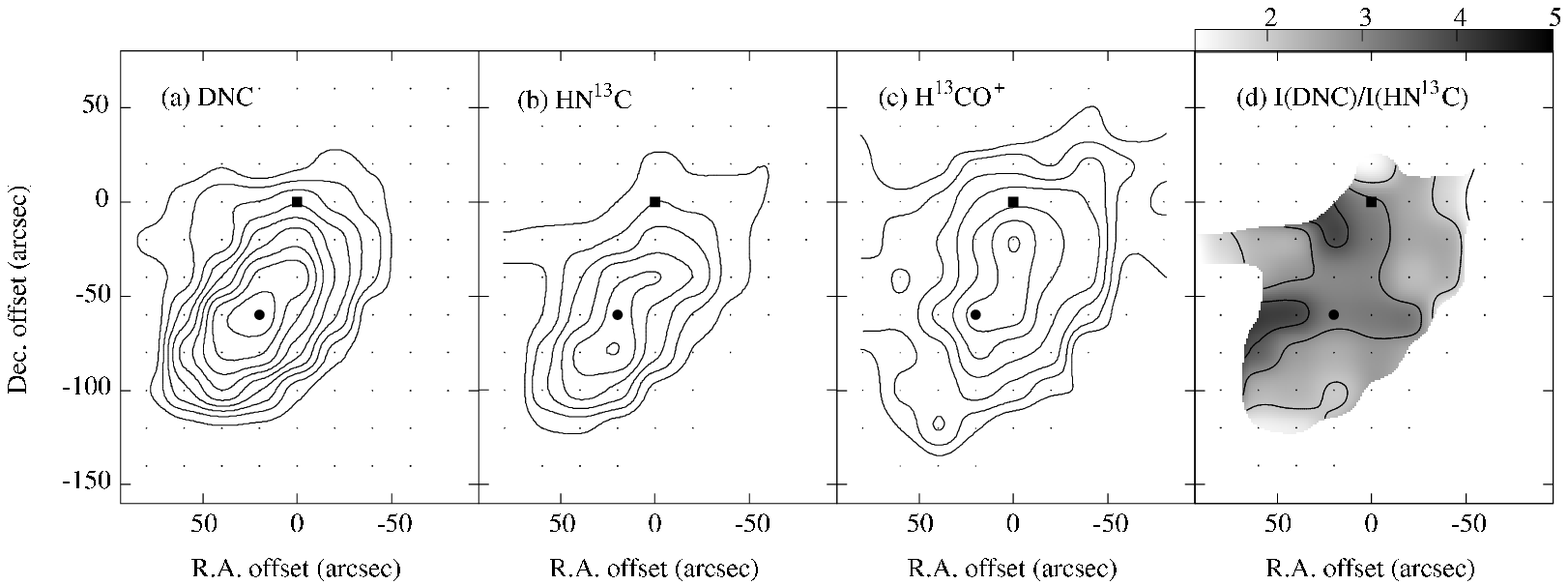}
\caption{Integrated intensity maps of L1544. 
Filled square and filled circle represent the (0\arcsec,0\arcsec) 
position and (+20\arcsec, -60\arcsec) position, respectively.  
(a) DNC ($J$=1-0). 
The velocity range of integration is from 6.2 to 8.5 km s$^{-1}$. 
The interval of the contours is 0.30 K km s$^{-1}$ and 
the lowest one is 0.60 K km s$^{-1}$. 
(b) HN$^{13}$C ($J$=1-0). 
The velocity range of integration is from 6.6 to 8.1 km s$^{-1}$. 
The interval of the contours is 0.16 K km s$^{-1}$ and 
the lowest one is 0.32 K km s$^{-1}$. 
(c) H$^{13}$CO$^{+}$ ($J$=1-0). 
The velocity range of integration is from 6.7 to 7.9 km s$^{-1}$. 
The interval of the contours is 0.12 K km s$^{-1}$ and 
the lowest one is 0.36 K km s$^{-1}$.  
(d) Integrated intensity ratio of DNC/HN$^{13}$C ($J$=1-0). 
Contour levels are 10, 30, 50, 70, and 90 \% of the peak value (4.18). 
\label{fig-l1544map} }
\end{figure}

\begin{figure}
\epsscale{1}
\plotone{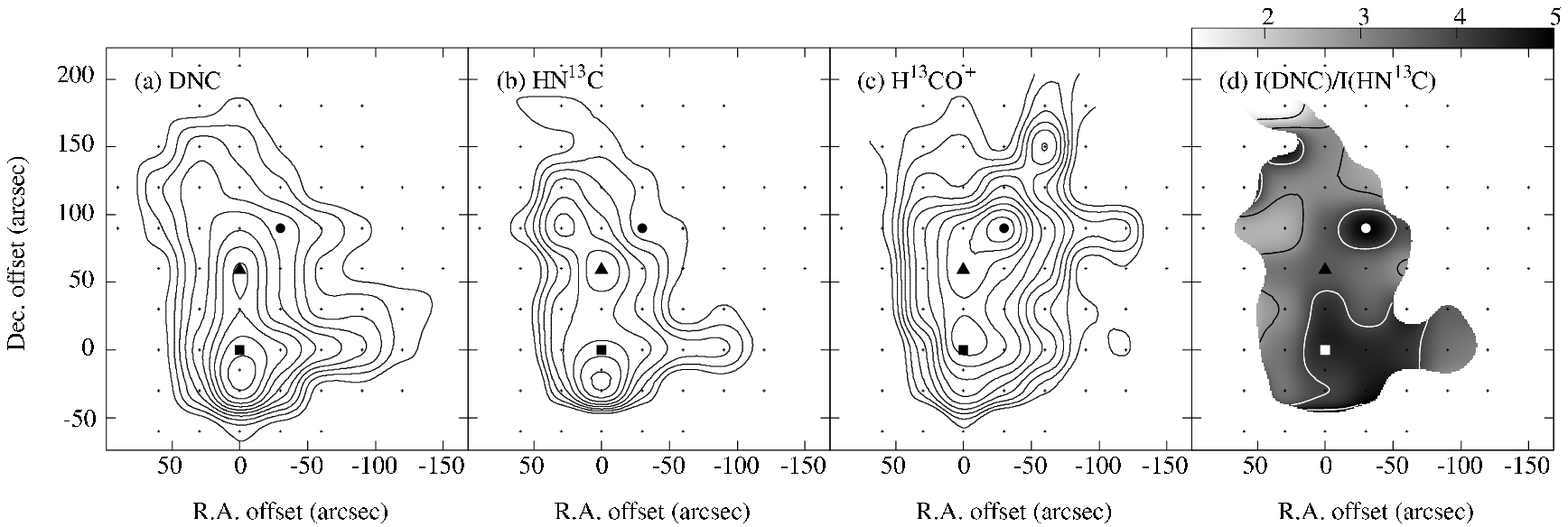}
\caption{Integrated intensity maps of L63. 
Filled square, filled circle, and filled triangle represent the 
(0\arcsec,0\arcsec) position, (-30\arcsec, +90\arcsec), 
and (0\arcsec, +60\arcsec) position, respectively.  
(a) DNC ($J$=1-0). 
The velocity range of integration is from 5.0 to 6.9 km s$^{-1}$. 
The interval of the contours is 0.24 K km s$^{-1}$ and 
the lowest one is 0.48 K km s$^{-1}$. 
(b) HN$^{13}$C ($J$=1-0). 
The velocity range of integration is from 5.4 to 6.5 km s$^{-1}$. 
The interval of the contours is 0.06 K km s$^{-1}$ and 
the lowest one is 0.24 K km s$^{-1}$. 
(c) H$^{13}$CO$^{+}$ ($J$=1-0). 
The velocity range of integration is from 5.4 to 6.5 km s$^{-1}$. 
The interval of the contours is 0.06 K km s$^{-1}$ and 
the lowest one is 0.36 K km s$^{-1}$.  
(d) Integrated intensity ratio of DNC/HN$^{13}$C ($J$=1-0). 
Contour levels are 10, 30, 50, 70, and 90 \% of the peak value (5.74). 
\label{fig-l63map} }
\end{figure}

The integrated intensity maps of DNC ($J$=1-0), HN$^{13}$C ($J$=1-0), 
and H$^{13}$CO$^{+}$ ($J$=1-0) along with the integrated intensity 
ratio maps of DNC/HN$^{13}$C ($J$=1-0) are shown in 
Figures \ref{fig-tmc1map}-\ref{fig-l63map}. 
As shown in Table \ref{tab-lvg}, 
the optical depths of the $J$=1-0 lines of DNC and HN$^{13}$C 
are less than 1.7. Therefore, the integrated intensity 
distributions of the $J$=1-0 lines approximately represent the column 
density distributions, and the integrated intensity ratio of 
DNC/HN$^{13}$C in the $J$=1-0 lines approximately 
represents the column density ratio of DNC/HN$^{13}$C. 
In fact, the integrated intensity ratios of DNC/HN$^{13}$C well 
correlate with the column density ratios, 
as shown in Table \ref{tab-dhratio}. 

In general, the distributions of DNC and HN$^{13}$C resemble 
that of N$_{2}$H$^{+}$ for L1512, L1544, and L63 
(Caselli et al. 2002c). 
On the other hand, the distributions of H$^{13}$CO$^{+}$ is 
more extended than those of DNC and HN$^{13}$C. 
These differences would reflect the chemical abundance variation 
within the core. Although these molecular lines trace 
similar density regions ($10^{5}$-$10^{6}$ cm$^{-3}$), 
DNC and HN$^{13}$C 
reside in an inner part of the core as N$_{2}$H$^{+}$ and NH$_{3}$ 
(Bergin \& Langer 1997; Caselli et al. 2002b). 

The distributions of DNC and HN$^{13}$C are also quite 
similar to each other. As a result, 
the variation of the integrated intensity ratio of DNC/HN$^{13}$C 
in each core is less than a factor of 2. 
We should emphasize that the variation of the DNC/HN$^{13}$C ratios 
from core to core, ranging from 0.50 to 7.3 (Paper I), 
is much larger than that within each core. 
Nevertheless, a small but systematic 
variation of the DNC/HN$^{13}$C ratios are found in 
TMC-1 and L63. These cores are larger in size than L1512 and L1544, 
and have multiple peak positions in the maps. 
First, we discuss these two cores. 

\subsection{TMC-1 and L63}

TMC-1 is one of the most well studied dark cloud cores, and 
various molecular lines have been detected with radio telescopes. 
The maps of the DNC, HN$^{13}$C, and H$^{13}$CO$^{+}$ lines 
in TMC-1 are shown in Figure \ref{fig-tmc1map}. 
As suggested in Paper I, a gradient can be seen in 
the DNC/HN$^{13}$C intensity ratio  along the ridge; 
the DNC/HN$^{13}$C intensity ratio around the ammonia 
peak in the northwest part of the ridge is systematically larger 
than that around the cyanopolyyne peak in the southeast part of 
the ridge. This is confirmed by our LVG analysis, as shown in 
Table \ref{tab-dhratio}. 
A similar trend can also be seen in other deuterated molecules 
such as DCO$^{+}$ 
(Gu\'elin, Langer, \& Wilson 1982; Saito et al. 2002), 
CH$_{2}$DCCH (Markwick, Millar, \& Charnley 2002), 
DC$_{3}$N (Howe et al. 1994), and C$_{3}$HD (Bell et al. 1988). 
Although the H$_{2}$ density is higher in the core C, 
the density gradient is not significant between 
the cyanoplyyne peak and the ammonia peak. 

As mentioned in Section 1, it is well known that the abundances 
of carbon-chain molecules show anti-correlation with those of 
NH$_{3}$, N$_{2}$H$^{+}$, and SO along the TMC-1 ridge 
(Hirahara et al. 1992; 1995). 
Thus, the DNC/HNC ratio well correlates with the NH$_{3}$/CCS ratio 
along the ridge. This correlation seems to be understood naturally 
in terms of chemical evolution; the southern part of the 
ridge is younger than the northern part (Paper I; Saito et al. 2002), 
although the other mechanism considering protostellar 
activities in the northern region is also presented 
(Markwick, Millar, \& Charnley 2000; 
Markwick, Charnley, \& Millar 2001)

A gradient in the ratio is marginally seen in L63. 
L63 is one of the cores with the highest deuterium 
fractionation in DNC (Paper I), 
DCO$^{+}$ (Butner et al. 1995), and NH$_{2}$D (Saito et al. 2000). 
The maps of the DNC, HN$^{13}$C, and H$^{13}$CO$^{+}$ lines 
in L63 are shown in Figure \ref{fig-l63map}. There are a few 
emission peaks in each map, and their positions are slightly 
different from one another. 
The overall distributions of DNC and HN$^{13}$C are similar to that 
of N$_{2}$H$^{+}$ (Caselli et al. 2002c). 
Millimeter and submillimeter continuum emissions were also 
observed by Ward-Thompson et al. (1994, 1999) and they  
presented the maps in the 1.3 mm and 800 $\mu$m emissions, although 
they only observed the southern part of the core. 
The DNC/HN$^{13}$C ratio is higher along the north-south ridge of 
the core and the (-30\arcsec, 90\arcsec) position, 
which corresponds to one of the H$^{13}$CO$^{+}$ peak. 
Especially, the (-30\arcsec, 90\arcsec) position is remarkable 
because the HN$^{13}$C line is only marginally detected. 
In this position, it is possible 
that the fractionation of DNC is as high as 
0.1-0.2. However, the physical and chemical conditions toward 
northern part of L63 are unclear in the present stage 
because of the lack of 
detailed observational data. 

\subsection{L1512 and L1544}

The maps of the DNC, HN$^{13}$C, and H$^{13}$CO$^{+}$ lines 
in L1512 and L1544 are shown in Figures \ref{fig-l1512map} and 
\ref{fig-l1544map}, respectively. 
In these cores, we cannot find apparent systematic 
gradients in the DNC/HN$^{13}$C ratio maps. 
Even when we obtained the column densities for two 
representative positions, (0\arcsec, 0\arcsec) and 
(20\arcsec, -60\arcsec), in L1544 by the LVG analysis, 
the DNC/HNC ratio is higher only by a factor of 1.5 toward the 
core center (20\arcsec, -60\arcsec) than toward the edge 
(0\arcsec, 0\arcsec). On the other hand, 
Caselli et al. (2002b) reported that the DCO$^{+}$/HCO$^{+}$ ratio 
in L1544 is enhanced almost by an order of magnitude from the edge 
to the center of the core. 
In addition, they also reported that 
the N$_{2}$D$^{+}$/N$_{2}$H$^{+}$ ratio is enhanced at 
the core center although the 
gradient is smaller than that of the DCO$^{+}$/HCO$^{+}$ ratio 
(Caselli et al. 2002b).
Therefore, our result is a contrast to theirs. As for L1512, 
no mapping observations 
of deuterated molecules has been reported as far as we know. 

It has been reported that the CO molecule is significantly depleted 
in the central part of L1544 and L1512 (Caselli et al. 2002b; 
Lee et al. 2003). 
Particularly, Caselli et al. (2002b) presented a map of the depletion 
factor of CO in L1544, indicating that the depletion factor increases 
at least by a factor of about 3 from the edge to the center of the 
core. A heavier depletion of CO would make the life time of 
H$_{2}$D$^{+}$ longer, resulting high H$_{2}$D$^{+}$/H$_{3}^{+}$ ratio 
(e.g. Roberts \& Millar 2000; Turner 2001). Therefore, the DNC/HNC 
ratio would also be enhanced through the deuteron transfer reaction 
from H$_{2}$D$^{+}$ followed by the electron recombination reaction. 
However, the present results suggest that the DNC/HNC ratio is not 
very sensitive to the depletion factor unlike DCO$^{+}$/HCO$^{+}$ 
and N$_{2}$D$^{+}$/N$_{2}$H$^{+}$ ratios. 

\section{Origin of the Variation of the DNC/HNC Ratio}

As discussed in Section 4, the DNC/HNC ratio is derived to be 
0.02-0.09 for the observed cores. These values are significantly 
higher than those predicted by pure gas-phase 
chemical models in cold (10 K) dark cloud cores, ranging 
from 0.015 to 0.03 (Howe \& Millar 1993; Roberts \& Millar 2000; 
Turner 2001). 
Hence the model including the effect 
of depletion of molecules onto grain surface is invoked 
to explain the observed ratio. According to Turner (2001), 
the DNC/HNC ratio increases by a factor of 2-2.5, as the depletion 
factor of C and O increases from 1 to 3. In order to 
confirm this effect further, we have investigated the spatial 
distribution of the DNC/HNC ratio in the L1544 core, 
where a positive gradient of the depletion factor 
from the edge to the center of the core is reported 
(Caselli et al. 2002b). 
However, no significant enhancement in the DNC/HNC ratio has been 
observed, although Caselli et al. (2002b) reported that 
the DCO$^{+}$/HCO$^{+}$ ratio is much enhanced toward 
the center of the core. 

One possibility for a lack of enhancement in the DNC/HNC ratio 
at the core center is a relatively small distribution of 
HNC. As noted in Section 5, the distribution of 
HNC is slightly smaller than that of HCO$^{+}$, which implies 
that DNC and HNC reside only in an inner part of the core 
than DCO$^{+}$ and HCO$^{+}$. If so, a variation of the 
depletion factor in such a restricted part might be small, 
which results in a relatively small variation of the DNC/HNC ratio. 
However, the large core-to-core variation of the DNC/HNC ratio 
cannot be explained in this way, unless the core-to-core variation 
of the depletion factor is introduced a priori. 

Another possibility is related to the time scale of the 
deuterium fractionation. The time scale for deuterium fractionation 
via the gas-phase reactions is roughly 
estimated to be 10$^{5}$ yr for the H$_{2}$ density of 
10$^{5}$ cm$^{-3}$, according to the chemical model calculations 
(Turner 2001). This would mainly be determined by the mean 
collision time of a molecule with the H$_{3}^{+}$ ion. 
This time scale is comparable to the dynamical 
time scale ($\sim$10$^{5}$ yr) 
and the depletion time scale ($\sim$10$^{5}$ yr). 
Therefore, there is a 
possibility that deuterium fractionation does not follow the 
change in depletion in a dynamically evolving core. 
This may be a reason why we observed no significant enhancement 
of the DNC/HNC ratio at the center of the core. The time scale 
can, of course, be different from species to species. According 
to Turner (2001), the time dependence of deuterium fractionation 
strongly depends on species. If the protonated species such as 
HCO$^{+}$ and N$_{2}$H$^{+}$ would have faster time scale for the 
deuterium fractionation, the DCO$^{+}$/HCO$^{+}$ ratio can be 
enhanced at the center of the core, as observed by 
Caselli et al. (2002b). 

The latter explanation can also account for the core-to-core 
variation of the DNC/HNC ratio. If a core contracts very slowly 
in comparison with the dynamical time scale, molecules are subject 
to heavy depletion. As a result, high deuterium fractionation would 
be expected. In fact, Aikawa et al. (2001) constructed a chemical 
model for a dynamically evolving core, and showed that 
the DCO$^{+}$/HCO$^{+}$ ratio would increase 
if the time scale of collapse is relatively longer than those of 
gas-phase chemistry and depletion of molecules onto grain surface. 
Probably the L63 core corresponds to this case. 
If the core is in a very early stage of collapse, 
deuterium fractionation has not well proceeded. 
Furthermore, molecules have not been depleted onto dust grains yet, 
which also gives an unfavorable condition for deuterium fractionation. 
Probably TMC-1 (CP) corresponds to this case, where the chemical 
age of the core seems to be young judging from the low 
NH$_{3}$/CCS ratio (Suzuki et al. 1992; Hirahara et al. 1992). 
In addition, the L1521E core also corresponds 
to this case, where an extremely low DNC/HNC ratio is reported 
(Hirota et al. 2002). From these considerations, the observed 
variation of the DNC/HNC ratio seems to originate from a 
combination of the three competing time-dependent processes; 
gas-phase deuterium fractionation, depletion of molecules onto 
grain surface, and dynamical evolution of a core 
(e.g. Aikawa et al. 2001). Therefore, 
the DNC/HNC ratio becomes higher for more chemically evolved 
cores, and shows a correlation with the NH$_{3}$/CCS ratio 
(Paper I). For quantitative discussions, systematic chemical 
model calculations of collapsing cores would be necessary. 
Such an effort has recently been carried out by 
Aikawa, Ohashi, \& Herbst (2003). 

\section{Summary}

In this paper, we report the results of 
mapping observations of the 
$J$=1-0 lines of DNC, HN$^{13}$C, and H$^{13}$CO$^{+}$ 
toward 4 nearby dark cloud cores; TMC-1, L1512, L1544, and L63. 
Along with the mapping observations, 
we have observed the $J$=2-1 lines of DNC and HN$^{13}$C 
toward selected positions, which enables us to 
determine the H$_{2}$ densities of the emission regions and 
the column densities of DNC and HN$^{13}$C by 
the statistical equilibrium calculations based on the LVG model. 
The main results of this paper are as follows: 

\begin{enumerate}
\item We calculated the column densities of DNC and HN$^{13}$C by 
the LVG calculations assuming that the DNC and HN$^{13}$C lines 
trace the same volume of gas. 
The DNC/HN$^{13}$C ratios range from 1.25 to 5.44, 
with a typical uncertainties of a factor of 2. 
The H$_{2}$ densities ranges from 1.4$\times$10$^{5}$ cm$^{-3}$ 
to 5.5$\times$10$^{5}$ cm$^{-3}$. 
\item Assuming the $^{12}$C/$^{13}$C ratio of 60, 
the DNC/HNC ratios are estimated to be 0.02-0.09. 
Some of the observed DNC/HN$^{13}$C ratios exceed values 
predicted by gas-phase chemical models in 
cold (10 K) dark cloud cores, 0.015-0.03. 
\item The distributions of DNC and HN$^{13}$C resemble that of 
N$_{2}$H$^{+}$ for L1512, L1544, and L63. 
On the other hand, the distributions of H$^{13}$CO$^{+}$ is 
more extended than those of DNC and HN$^{13}$C. 
These difference would reflect chemical abundance variation 
within the core; 
DNC and HN$^{13}$C would reside in an inner part of the core 
as well as the other nitrogen-bearing molecules such as 
N$_{2}$H$^{+}$ and NH$_{3}$ while H$^{13}$CO$^{+}$ in 
an outer envelope. 
\item The distributions of DNC and HN$^{13}$C are quite 
similar to each other, and the variation of the DNC/HN$^{13}$C 
ratio within each core is less than a factor of 2. 
The variation of the DNC/HN$^{13}$C ratios from core to core, 
ranging an order of magnitude (0.50-7.3), is rather remarkable than 
those within each core. 
\item While the systematic variation of the DNC/HN$^{13}$C ratio 
can be seen in TMC-1 and L63, a similar gradient cannot 
be found in L1512 and L1544. 
Particularly, the DNC/HNC ratio does not show enhancement at the 
center of the core of L1544 unlike the DCO$^{+}$/HCO$^{+}$ ratio. 
This indicates that the DNC/HNC ratio 
is not very sensitive to depletion factor. 
Since the time scale of gas-phase deuterium fractionation, depletion of 
molecules onto grain surface, and dynamical evolution of a cloud 
core are comparable 
to one another, a combination of these competing processes 
would be responsible for the variation of the 
DNC/HNC ratio. The DNC/HNC ratio tends to be higher for more 
chemically evolved cores.  
\end{enumerate}

\acknowledgements

We are grateful to the staff of Nobeyama Radio Observatory and 
the NRAO 12 m telescope for their assistance in observations. 
We are also grateful to Shuji Saito of Fukui University and 
Yuri Aikawa of Kobe University for valuable discussions. 
TH thanks to the Japan Society for the Promotion of Science 
for the financial support. 
This study is partly supported by Research Aid of Inoue Foundation 
for Science and Grant-in-Aid from Ministry of 
Education, Science, Sports and Culture of Japan 
(Nos. 07CE2002, 14204013).

{}

\newpage

\begin{deluxetable}{lclcc}
\tabletypesize{\scriptsize}
\tablenum{1}
\tablewidth{0pt}
\tablecaption{Observed lines 
\label{tab-obsedf}}
\tablehead{
\colhead{Molecule} & \colhead{Transition} &
  \colhead{$\nu$ (MHz)} &   \colhead{$\mu$ (D)} & 
  \colhead{$S_{ul}$\tablenotemark{a}}
}
\startdata
DNC            & $J$=1-0 & 76305.727 & 3.05  &  1.00  \\
DNC            & $J$=2-1 & 152609.774 & 3.05  &  2.00  \\
HN$^{13}$C       & $J$=1-0 & 87090.850 & 3.05 & 1.00  \\
HN$^{13}$C       & $J$=2-1 & 174179.408 & 3.05 & 2.00  \\
H$^{13}$CO$^{+}$ & $J$=1-0 & 86754.330 & 4.07 & 1.00 \\
\enddata
\tablenotetext{ a}{ Intrinsic line strength}
\end{deluxetable}

\begin{deluxetable}{lcccc}
\tabletypesize{\scriptsize}
\tablenum{2}
\tablewidth{0pt}
\tablecaption{Summary of mapping observations with 
the NRO 45 m telescope
\label{tab-mapobs}}
\tablehead{
\colhead{Source} & \colhead{$\alpha$(1950)} &
  \colhead{$\delta$(1950)}  & \colhead{Grid} & \colhead{} \\
\colhead{Name} & \colhead{ (  h  m  s)} &
  \colhead{(  $^{\circ}$ \hspace{0.3em} \arcmin  \hspace{0.3em} \arcsec)} &
  \colhead{(arcsec)} & \colhead{Date} \\}
\startdata
TMC-1   & 04 38 38.6 & \ 25 35 45 & 60 & Apr 1997, Feb 1998 \\
L1512   & 05 00 54.4 & \ 32 39 00 & 20 & May 1999 \\
L1544   & 05 01 11.1 & \ 25 07 40 & 20 & Apr 1997, Feb 1998 \\
L63     & 16 47 21.0 & -18 01 00  & 30 & Feb 1999 \\
\enddata
\end{deluxetable}

\begin{deluxetable}{lclccccc}
\tabletypesize{\scriptsize}
\tablenum{3}
\tablewidth{0pt}
\tablecaption{Line parameters for the DNC and HN$^{13}$C lines 
\label{tab-obs}}
\tablehead{
\colhead{Source} & \colhead{} & \colhead{} & 
  \colhead{$T_{mb}$} & \colhead{$v_{lsr}$} &

  \colhead{$\Delta v$} & \colhead{$T_{rms}$} & \colhead{}    \\
\colhead{Name} & \colhead{Position} & 
  \colhead{Molecule} & \colhead{ (K)} & \colhead{(km s$^{-1}$)} &
  \colhead{(km s$^{-1}$)} &  \colhead{(K)} & \colhead{References} \\}
\startdata
TMC-1  & (CP\tablenotemark{a}) 
 & DNC(1-0) & 1.42(0.16) & 5.89(0.06) & 1.16(0.16) & 0.16 & 1  \\ 
 & & HN$^{13}$C(1-0) & 1.12(0.17) & 5.94(0.06) & 0.75(0.14) & 0.13 & 1 \\ 
 & & DNC(2-1) & 0.43(0.06) & 5.92(0.06) & 0.83(0.13) & 0.04 &  \\ 
 & & HN$^{13}$C(2-1) & 0.41(0.08) & 5.82(0.06) & 0.54(0.13) & 0.05 &  \\ 
           &  (NH$_{3}$\tablenotemark{a}) 
 & DNC(1-0) & 1.76(0.17) & 5.90(0.05) & 1.12(0.13) & 0.17 & 1 \\ 
 & & HN$^{13}$C(1-0) & 1.07(0.21) & 5.96(0.07) & 0.73(0.18) & 0.14 & 1 \\ 
 & & DNC(2-1) & 0.78(0.08) & 5.99(0.05) & 0.86(0.11) & 0.05 &  \\ 
 & & HN$^{13}$C(2-1) & 0.39(0.13) & 6.03(0.09) & 0.52(0.21) & 0.09 &  \\ 
            &   (core C\tablenotemark{a})
 & DNC(1-0) & 1.75(0.25) & 5.89(0.07) & 1.01(0.17) & 0.21 &  \\ 
 & & HN$^{13}$C(1-0) & 1.32(0.21) & 6.00(0.05) & 0.64(0.12) & 0.16 &  \\ 
 & & DNC(2-1) & 0.88(0.09) & 5.98(0.04) & 0.79(0.10) & 0.06 &  \\ 
 & & HN$^{13}$C(2-1) & 0.79(0.11) & 5.85(0.04) & 0.50(0.09) & 0.07 &  \\ 
L1512 &  (0\arcsec, 0\arcsec) 
 & DNC(1-0) & 1.28(0.12) & 7.13(0.04) & 0.82(0.09) & 0.06 &1  \\ 
 & & HN$^{13}$C(1-0) & 1.25(0.10) & 7.19(0.02) & 0.51(0.05) & 0.06 & 1 \\ 
 & & DNC(2-1) & 0.68(0.06) & 7.18(0.03) & 0.78(0.08) & 0.05 &  \\ 
 & & HN$^{13}$C(2-1) & 0.63(0.12) & 7.10(0.04) & 0.41(0.09) & 0.06 &  \\ 
           & (-40\arcsec, 0\arcsec) 
 & DNC(1-0) & 1.46(0.25) & 7.12(0.08) & 0.92(0.19) & 0.16 &  \\ 
 & & HN$^{13}$C(1-0) & 1.13(0.22) & 7.22(0.05) & 0.51(0.12) & 0.14 &  \\ 
 & & DNC(2-1) & 0.64(0.13) & 7.12(0.08) & 0.74(0.19) & 0.08 &  \\ 
 & & HN$^{13}$C(2-1) & 0.50(0.31) & 7.03(0.00) & 0.25(0.20) & 0.15 &  \\ 
L1544 &   (0\arcsec, 0\arcsec) 
 & DNC(1-0) & 1.33(0.17) & 7.05(0.07) & 1.18(0.19) & 0.15 & 1 \\ 
 & & HN$^{13}$C(1-0) & 0.86(0.15) & 7.18(0.07) & 0.80(0.17) & 0.14 & 1 \\ 

 & & DNC(2-1) & 0.67(0.07) & 7.15(0.04) & 0.82(0.10) & 0.06 &  \\ 
 & & HN$^{13}$C(2-1) & 0.40(0.13) & 7.06(0.11) & 0.64(0.26) & 0.08 &  \\ 
             &  (20\arcsec, -60\arcsec) 
 & DNC(1-0) & 2.91(0.32) & 7.21(0.06) & 1.09(0.15) & 0.25 &  \\ 
 & & HN$^{13}$C(1-0) & 1.71(0.27) & 7.16(0.04) & 0.53(0.10) & 0.16 &  \\ 
 & & DNC(2-1) & 1.69(0.10) & 7.17(0.03) & 0.93(0.06) & 0.07 &  \\ 
 & & HN$^{13}$C(2-1) & 0.92(0.15) & 7.15(0.06) & 0.67(0.13) & 0.10 &  \\ 
L63 & (0\arcsec, 0\arcsec) 
 & DNC(1-0) & 2.54(0.32) & 5.74(0.06) & 0.88(0.14) & 0.17 & 1 \\ 
 & & HN$^{13}$C(1-0) & 0.92(0.36) & 5.86(0.09) & 0.46(0.22) & 0.17 &1  \\ 
 & & DNC(2-1) & 1.27(0.07) & 5.79(0.02) & 0.78(0.05) & 0.05 &  \\ 
 & & HN$^{13}$C(2-1) & 0.49(0.18) & 5.68(0.07) & 0.38(0.17) & 0.09 &  \\ 
        & (-30\arcsec, 90\arcsec)
 & DNC(1-0) & 1.38(0.27) & 5.75(0.08) & 0.86(0.20) & 0.21 &  \\ 
 & & HN$^{13}$C(1-0) & \nodata & \nodata & \nodata & 0.17 &  \\ 
 & & DNC(2-1) & 0.47(0.05) & 5.75(0.04) & 0.77(0.10) & 0.04 &  \\ 
 & & HN$^{13}$C(2-1) & \nodata & \nodata & \nodata & 0.07 &  \\ 
       & (0\arcsec, 60\arcsec) 
 & DNC(1-0) & 2.43(0.24) & 5.80(0.04) & 0.88(0.10) & 0.18 &  \\ 
 & & HN$^{13}$C(1-0) & 0.96(0.19) & 5.88(0.07) & 0.69(0.16) & 0.14 &  \\ 
 & & DNC(2-1) & 0.80(0.10) & 5.79(0.05) & 0.76(0.11) & 0.07 &  \\ 
 & & HN$^{13}$C(2-1) & \nodata & \nodata & \nodata & 0.14 &  \\ 
\enddata
\tablerefs {1:Hirota et al. 2001}
\tablenotetext{a}{CP and NH$_{3}$ correspond to 
the cyanopolyyne peak and NH$_{3}$ peak, respectively. 
Core C corresponds to the position labeled by 
Hirahara et al. (1992). The coordinates (B1950) 
and offsets from the reference position (CP peak) of 
NH$_{3}$ peak and core C are as follows; 
%NH$_{3}$: 04 38 14.5, 25 42 30 
NH$_{3}$: 04$^{h}$38$^{m}$14$^{s}$.5, 
25$^{\circ}$42\arcmin30\arcsec 
(-5\arcmin26\arcsec, 6\arcmin45\arcsec); 
%core C: 04 38 29.2, 25 39 17 
core C: 04$^{h}$38$^{m}$29$^{s}$.2, 
25$^{\circ}$39\arcmin17\arcsec 
(-2\arcmin07\arcsec, 3\arcmin32\arcsec).}
\tablecomments{The numbers in parenthesis represent 
three times the standard deviation in the Gaussian fit (3$\sigma$).}
\end{deluxetable}

\begin{deluxetable}{lclccccccc}
\tabletypesize{\scriptsize}
\tablenum{4}
\tablewidth{0pt}
\tablecaption{Results of the LVG calculations 
\label{tab-lvg}}
\tablehead{
\colhead{Source} & \colhead{} 
  & \colhead{} & \multicolumn{2}{c}{$J$=1-0} & \colhead{} 
  & \multicolumn{2}{c}{$J$=2-1} & \colhead{$n$(H$_{2}$)} 
  & \colhead{$N$} \\
\cline{4-5} \cline{7-8}
\colhead{Name}   & \colhead{Position} 
  & \colhead{Molecule} & \colhead{$\tau$} & \colhead{$T_{ex}$(K)} 
  & \colhead{} & \colhead{$\tau$} & \colhead{$T_{ex}$(K)} 
  & \colhead{(10$^{5}$ cm$^{-3}$)} & \colhead{(10$^{12}$ cm$^{-2}$)} \\}
\startdata
TMC-1 & (CP) 
    & DNC &  0.8   &  4.9  & &  1.2   &  3.7   &  1.8\tablenotemark{a}   &  3.46 \\
  & & HN$^{13}$C &  1.3   &  4.5  & &  1.5   &  3.6   &  1.8   &  2.77 \\
      & (NH$_{3}$) 
   & DNC &  1.6   &  5.1 &  &  2.5   &  4.0     &  1.4  &  7.56  \\

 & & HN$^{13}$C &  1.5   &  4.3  & &  1.7   &  3.5     &  1.4  &  3.03  \\
      & (core C) 
     & DNC &  0.4   &  8.0 &  &  1.0   &  4.8     &  5.5  &  2.70  \\
 & & HN$^{13}$C &  0.6   &  6.6 &  &  1.1   &  4.4     &  5.5  &  1.74  \\
L1512 & (0\arcsec, 0\arcsec) 
   & DNC &  0.4   &  7.2 &  &  0.9   &  4.4     &  4.6  &  2.08  \\
 & & HN$^{13}$C &  0.6   &  5.9  & &  1.0   &  4.2     &  4.6  &  1.26  \\
      & (-40\arcsec, 0\arcsec) 
   & DNC &  0.9   &  5.2 &  &  1.6   &  3.9   &  2.0  &  3.89  \\
 & & HN$^{13}$C &  1.7   &  4.6 &  &  1.4   &  3.7     &  2.0  &  1.54  \\
L1544 & (0\arcsec, 0\arcsec) 
   & DNC &  0.5   &  6.4 &  &  1.0   &  4.3     &  3.5  &  3.02  \\
 & & HN$^{13}$C &  0.5   &  5.2 &  &  0.7   &  3.9     &  3.5  &  1.51  \\
      & (20\arcsec, -60\arcsec) 
   & DNC &  0.8   &  7.8 &  &  2.1   &  5.3     &  4.8  &  7.31  \\
 & & HN$^{13}$C &  0.8   &  6.4 &  &  1.4   &  4.5     &  4.8  &  2.44  \\
L63 &  (0\arcsec, 0\arcsec) 
   & DNC &  1.4   &  6.2 &  &  2.7   &  4.6     &  2.4  &  7.29 \\
 & & HN$^{13}$C &  0.9   &  4.7 &  &  1.1   &  3.7     &  2.4  &  1.34  \\
      & (-30\arcsec, 90\arcsec)
   & DNC &  0.6   &  5.4  & &  1.0   &  3.9   &  2.4\tablenotemark{b}  &  2.35 \\
 & & HN$^{13}$C & \nodata & \nodata & & \nodata & \nodata &  2.4\tablenotemark{b}   & $<$0.24 \\
       & (0\arcsec, 60\arcsec) 
   & DNC &  1.0   &  5.9  & &  2.0   &  4.3   &  2.4\tablenotemark{b}   &  4.92 \\
  & & HN$^{13}$C & 0.7 & 4.6 & &  \nodata & \nodata & 2.4\tablenotemark{b} & 1.80  \\
\enddata
\tablenotetext{a}{H$_{2}$ density derived from the HN$^{13}$C data 
toward TMC-1 (CP). }
\tablenotetext{b}{H$_{2}$ density derived from 
the DNC and HN$^{13}$C data toward L63 (0\arcsec, 0\arcsec). }
\end{deluxetable}

\begin{deluxetable}{lclcccc}
\tabletypesize{\scriptsize}
\tablenum{5}
\tablewidth{0pt}
\tablecaption{DNC/HN$^{13}$C ratios in the observed cores
\label{tab-dhratio}}
\tablehead{
\colhead{Source}
  & \colhead{} & \multicolumn{2}{c}{$I$(DNC)/$I$(HN$^{13}$C)} & \colhead{} 
  & \multicolumn{2}{c}{$N$(DNC)/$N$(HN$^{13}$C)} \\
\cline{3-4} \cline{6-7}
\colhead{Name}   & \colhead{Position} 
  & \colhead{$J$=1-0} & \colhead{$J$=2-1} 
  & \colhead{} & \colhead{Present}  & \colhead{Paper I} 
  \\}
\startdata
TMC-1  &  (CP)    &  
  1.96(0.37)  &   1.61(0.27)  &   & 1.25 &  1.70(0.11)  \\
       &  (NH$_{3}$)    &  
  2.52(0.50)  &   3.31(0.92)  &   &  2.50 & 2.5(0.2)  \\
       &  (core C)    &  
  2.09(0.44)  &   1.76(0.23)  &   &  1.55 & \nodata   \\
L1512  &  (0\arcsec, 0\arcsec)     &  
  1.65(0.15)  &   2.05(0.31)  &    &  1.65 & 1.34(0.04)   \\
       &  (-40\arcsec, 0\arcsec)     &  
  2.33(0.52)  &   3.79(2.00)  &   &  2.53 & \nodata   \\
L1544  &  (0\arcsec, 0\arcsec)     &  
  2.28(0.52)  &   2.15(0.51)  &   &  2.00 & 2.0(0.2)   \\
       &  (20\arcsec, -60\arcsec)     &  
  3.50(0.58)  &   2.55(0.30)  &  &  3.00 & \nodata  \\
L63  &   (0\arcsec, 0\arcsec)     &  
  5.28(1.60)  &   5.32(2.44)  &   &  5.44 & 7.0(0.7)   \\
       &  (-30\arcsec, 90\arcsec)    &  
     $>$8.87 & $>$5.75 &    & $>$9.79 &  \nodata    \\
        &  (0\arcsec, 60\arcsec)     &  
  3.23(0.68)  &   $>$4.94  &    & 2.73 &  \nodata  \\
\enddata
\tablecomments{The numbers in parenthesis represent the errors 
estimated from the rms noise of the observed spectra(3$\sigma$).}
\end{deluxetable}

%%%%%%%%%%%%%%%%%%%%%%%%%%%%%%%%%%%%%%%%%%%%%%%%%%%%%%%%%%%%%%%%
%Figures
%
\newpage



\begin{thebibliography}{}
\bibitem[aik01]{aik01}
Aikawa, Y., Ohashi, N., Inutsuka, S., Herbst, E., \& Takakuwa, S. 
  2001, ApJ, 552, 639
\bibitem[aik03]{aik03}
Aikawa, Y., Ohashi, N., \& Herbst, E. 2003, ApJ, in press
\bibitem[bel88]{bel88}
 Bell, M. B., Avery, L. W., Matthews, H. E., Feldman, P. A., 
  Watson, J. K. G., Madden, S. C., \& Irvine, W. M. 1988, 
  ApJ, 326, 924
\bibitem[ben89]{ben89}
 Benson, P. J., \& Myers, P. C. 1989, ApJS,
  71, 89
\bibitem[ber97]{ber97}
 Bergin, E. A. \& Langer, W. D. 1997, ApJ, 486, 316
\bibitem[bla76]{bla76}
 Blackman, G. L., Brown, R. D., Godfrey, P. D., \& Gunn, H. I.
  1976, Nature, 261, 395
\bibitem[but95]{but95}
 Butner, H. M., Lada, E. A., \& Loren, R. B. 1995, 
  ApJ, 448, 207
\bibitem[cas02a]{cas02a}
 Caselli, P., Walmsley, C. M., Zucconi, A., Tafalla, M., 
  Dore, L., \& Myers, P. C. 2002a, ApJ, 565, 331
\bibitem[cas02b]{cas02b}
 Caselli, P., Walmsley, C. M., Zucconi, A., Tafalla, M., 
  Dore, L., \& Myers, P. C. 2002b, ApJ, 565, 344
\bibitem[cas02c]{cas02c}
 Caselli, P., Benson, P. J., Myers, P. C., \& Tafalla, M. 
 2002c, ApJ, 572, 238
\bibitem[fre79]{fre79}
 Frerking, M. A., Langer, W. D., \& Wilson, R. W. 
  1979, ApJ, 232, L65
\bibitem[ger87]{ger87}
Gerin, M., et al. 1987, A\&A, 173, L1
\bibitem[gol74]{gol74}
 Goldreich, P., \& Kwan, J. 1974, ApJ, 189, 441
\bibitem[gre74]{gre74}
 Green, S., \& Thaddeus, P. 1974, ApJ, 191, 653
\bibitem[gue82]{gue82}
 Gu\'elin, M., Langer, W. D., \& Wilson, R. W. 
  1982, A\&A, 107, 107
\bibitem[hir92]{hir92}
 Hirahara, Y., et al. 1992, ApJ, 394, 539
\bibitem[hir95]{hir95}
 Hirahara, Y., et al. 1995, PASJ, 47, 845
\bibitem[hir01]{hir01}
 Hirota, T., Ikeda, M., \& Yamamoto, S. 2001, ApJ, 547, 814 (Paper I)
\bibitem[hir02]{hir02}
  Hirota, T., Ito, T., \& Yamamoto, S. 2002, ApJ, 565, 359
\bibitem[hir98]{hir98}
 Hirota, T., Yamamoto, S., Mikami, H., \& 
  Ohishi, M. 1998, ApJ, 503, 717
\bibitem[how93]{how93}
 Howe, D. A., \& Millar, T. J. 1993, MNRAS, 262, 868
\bibitem[how94]{how94}
 Howe, D. A., Millar, T. J., Schilke, P., \& Walmsley, C. M.
  1994, MNRAS, 267, 59
\bibitem[ike02]{ike02}
 Ikeda, M., Hirota, T., \& Yamamoto, S. 2002, ApJ, 575, 250
\bibitem[lan93]{lan93}
 Langer, W. D., \& Penzias, A. A. 1993, ApJ, 
  408, 539
\bibitem[lee03]{lee03}
  Lee, J.-E., Evans N. J., II, Shirley, Y. L., and Tatematsu, K. 
  2003, ApJ, 583, 789
\bibitem[man88]{man88}
 Mangum, J. G., Rood, R. T., Wadiak, E. J., \& 
  Wilson, T. L. 1988, ApJ, 334, 182
\bibitem[mar01]{mar01}
  Markwick, A. J., Charnley, S. B., \& Millar, T. J. 2001, 
  A\&A, 376, 1054
\bibitem[mar00]{mar00}
  Markwick, A. J., Millar, T. J., \& Charnley, S. B. 2000, 
  ApJ, 535, 256
\bibitem[mar02]{mar02}
  Markwick, A. J., Millar, T. J., \& Charnley, S. B. 2002, 
  A\&A, 381, 560
\bibitem[mil89]{mil89}
  Millar, T. J., Bennett, A., \& Herbst, E. 1989, ApJ, 340, 906
\bibitem[rob00]{rob00}
  Roberts, H. \& Millar, T. J. 2000, A\&A, 361, 388
\bibitem[sai00]{sai00}
 Saito, S., Ozeki, H., Ohishi, M., \& Yamamoto, S. 2000, 
  ApJ, 535, 227 
\bibitem[sai02]{sai02}
 Saito, S., Aikawa, Y., Herbst, E., Ohishi, M., Hirota, T., 
  Yamamoto, S., \& Kaifu, N. 2002, ApJ, 569, 836 
\bibitem[sha01]{sha01}
Shah, R. Y., \& Wootten, A. 2001, ApJ, 554, 933
\bibitem[suz92]{suz92}
 Suzuki, H., Yamamoto, S., Ohishi, M., Kaifu, N., Ishikawa, S., 
  Hirahara, Y., \& Takano, S. 1992, ApJ, 392, 551
\bibitem[taf98]{taf98}
 Tafalla, M., Mardones, D., Myers, P. C., Caselli, P., 
  Bachiller, R., \& Benson, P. J. 1998, ApJ, 504, 900
\bibitem[tur01]{tur01}
  Turner, B. E. 2001, ApJS, 136, 579
\bibitem[war94]{war94}
 Ward-Thompson, D., Scott, P. F., Hills, R. E., \& 
  Andr\'e, P. 1994, MNRAS, 268, 276
\bibitem[war99]{war99}
 Ward-Thompson, D., Motte, F., \& Andr\'e, P. 1999, 
  MNRAS, 305, 143

\bibitem[wil94]{wil94}
 Wilson, T. L. \& Rood, R. T. 1994, ARAA, 32, 191
\end{thebibliography}
\end{document}